\documentclass[conference]{IEEEtran}
\IEEEoverridecommandlockouts
\usepackage{amsmath,amssymb,amsfonts}
\usepackage{algorithmic}
\usepackage{graphicx}
\usepackage{subcaption}
\usepackage{amssymb}
\usepackage{pifont}
\usepackage{textcomp}
\usepackage{booktabs}
\usepackage[pscoord]{eso-pic}
\usepackage[a4paper, total={184mm,238mm}]{geometry}
\usepackage{xcolor}
\usepackage{float}
\usepackage{siunitx}
\usepackage{color,soul}
\usepackage[linesnumbered,lined,commentsnumbered]{algorithm2e}
\def\BibTeX{{\rm B\kern-.05em{\sc i\kern-.025em b}\kern-.08em
    T\kern-.1667em\lower.7ex\hbox{E}\kern-.125emX}}

\usepackage[newfloat,frozencache,cachedir=.]{minted}
\setminted[python]{breaklines, 
xleftmargin=2em, 
xrightmargin=1em, 
linenos=true, 
fontsize=\footnotesize}

\usepackage{caption}

\newenvironment{code}{\captionsetup{type=listing}
\medskip%
    \par%
    \noindent%
    }%
    {%
    \smallskip\par%
    }
\SetupFloatingEnvironment{listing}{name=Source Code}

\usepackage{hyperref}

\begin{document}
\title{Python Framework for Modular and Parametric SPICE Netlists Generation
}

\author{\IEEEauthorblockN{Sergio Vinagrero Gutiérrez, Giorgio Di Natale, Elena-Ioana Vatajelu}\\
\IEEEauthorblockA{\IEEEauthorblockA{\small{Univ. Grenoble Alpes, CNRS, Grenoble INP*, TIMA, 38000 Grenoble, France}}}
\IEEEauthorblockA{\small{\{sergio.vinagrero-gutierrez,giorgio.di-natale,ioana.vatajelu\}@univ-grenoble-alpes.fr}}
}

\maketitle

\begin{abstract}
Due to the complex specifications of current electronic systems, design decisions need to be explored automatically. However, the exploration process is a complex task given the plethora of design choices such as the selection of components, number of components, operating modes of each of the components, connections between the components and variety of ways in which the same functionality can be implemented. To tackle these issues, scripts are used generate designs based on high abstract constructions. Still, this approach is usually ad-hoc and platform dependent, making the whole procedure hardly reusable, scalable and versatile. We propose a generic, open-source framework tackling rapid design exploration for the generation of modular and parametric electronic designs that is able to work on any major simulator.

\let\thefootnote\relax\footnotetext{\textsuperscript{*}Institut National Polytechnique Grenoble Alpes}
\end{abstract}


\section{Introduction}

Design complexity, ultra-low-power requirements, reliability, robustness and security are becoming increasingly important concerns when designing electronic systems. Due to the increasing complexity of analog circuits, it is more difficult to design and assess their performance. Moreover, the aggressive scaling of CMOS technology makes the process of testing the same design under different technologies very tedious, as normally the process has to be repeated for every technology node. What is more, several issues must be considered at design time such as fabrication-induced variability, technology-dependent defects, extreme operating/environmental conditions, stochastic behaviours, aging, and possible perturbations (noise, radiations, malicious attacks). All these factors make the verification and testing of each circuit an arduous process. 

To explore the behaviour of an electrical circuit under different designs and conditions, multiple iterations and simulations need to be performed under the desired environment. The interdependencies of large and complex circuits can quickly become a significant challenge due to the extensive amount of choices at play. Design space exploration (DSE) examines the different possibilities and design options within the allowed design space considering the constraints and requirements in order to full fill the specified performance goals. DSE normally involves the use of tools as well as high-level abstract models of the system, to automate and streamline the exploration process since the design space is too large to be explored by hand. There is an interest in the industry to accelerate this process and reduce the time between iteration cycles. In the last decades there have been huge advances in computer aided design (CAD) and electronic design automation (EDA) to provide designers powerful tools to perform complex designs and characterisation. 

The idiosyncrasies of some technologies are very well understood and can be translated to higher levels of abstraction. However, with the present issues faced by today's designs, electrical-level simulations are unavoidable since they allow designers to accurately model and understand the behaviour of the target system. They are a crucial pillar of Analog and Mixed Signal design space exploration, simulation of circuit under the presence of perturbations and research of novel computation paradigms. But unlike digital circuits, where the low-level phases of the design process are automated using fairly standard methodologies, synthesis and layout of analog circuits is still carried out manually or through some sort of ad-hoc automated automated solution.

In this paper we show a Python framework~\cite{nimphel} for the generation of modular and reusable electronic designs through the use of powerful manipulation primitives. This purpose of this framework is twofold: (i) to provide tools to create electrical components whose characteristics can be expressed trough dynamic models or defined by logical ruls and (ii) to provide powerful manipulation primitives to quickly create complex arrangements of components in a simple fashion. The designs created are modular and reusable and can be serialised an even exported to work with any major available simulator.

This paper is organised as follows: the current state of the art is summarised in section~\ref{sec:state_art}, followed by the motivation for this project in section~\ref{sec:motivation}; In section~\ref{sec:description} the framework are described in detail and some use cases are provided in section~\ref{sec:example}. Conclusions are extracted in section~\ref{sec:conclusions}

\section{State of the art}\label{sec:state_art}

There are currently a plethora of tools available that tackle design space exploration. Tools like~\cite{chisel} and~\cite{pymtl3} provide frameworks with a high abstraction level that are able to compile a high level language, like Scala and Python, into fully functional Verilog code for hardware description. In this way, circuit designers have the expressiveness and power of a programming language in order to quickly create reusable circuits. These tools target Register Transfer Level (RTL) and thus are not very well suited for analog and mixed signal designs.

PySpice~\cite{PySpice} is an utility to generate SPICE netlists and launch simulations by embedding the design and the simulator configuration under the same language, which facilitates the whole design iteration process. However the simulator is limited to NGspice and Xyce and the netlists can only be exported for PCB designs. Skidl~\cite{Skidl} is a layer built on top of PySpice that attempts to to facilitate the process of connecting different components. Our framework seeks to provide designs for any available simulator as well as powerful tools to create complex connections and reusable components.

Alongside these tools, there are projects that provide automatic layout generation mechanism. One of the most famous known tools in this category is Magic~\cite{magic}. Magic is an interactive software for creating and modifying very large ncale integration (VLSI) circuit layouts. Its most important feature is the creation of a layout and \textit{plowing} it to scale it for different technology nodes. The ease of use of this utility comes with a penalty of 5 to 10\% increase in area usage. Other tools found online like LibreCell~\cite{LibreCell} try to reduce this tradeoff by reducing the fan of possibilities that are provided to the user. Lower level tools like GDSTK~\cite{gdstk} and GDSFactory~\cite{gdsfactory} enable the creation and manipulation of GDSII and OASIS files, which are the standard file format for foundries to specify circuit layouts. These tools can be used as the basis of a much more complete software that is able to generate the layout based on a circuit definition.

AIDA~\cite{aida} is a tool that AIDA tackles analog IC sizing and layout. It provides powerful utilities to perform parametric analysis, where the under laying parameters and properties of a circuit can be generated and swapped in place before every simulation cycle. However, the user needs to have generated the design beforehand, which does not solve the issue of design exploration.

There are complete projects like OpenRAM~\cite{OpenRAM} that provide a Python framework to create the layout, netlists, timing and power models, placement and routing models to use SRAMs in ASIC design. This tools provides an easy interface to configure the characteristics of the SRAM. This is a very powerful tool but is limited to SRAMs and a selected number of technology nodes.

Lastly, there are other projects found in the literature like~\cite{automated_netlist,youssef2011python} that provide tools that are crafted for an specific problem in mind. But as it has stated before, this tools are ad-hoc and provide almost no code re-usability.

\section{Motivation}\label{sec:motivation}

The main objective of this tool is to provide a framework that enables users to perform quick design space exploration and parametrization of electronic designs. A high abstract interface is provided in order to create modular and reusable components, that can be seaming less parameterised in order to provide users a general overview of the design under different design constraints and environments.

The advantage of using a programming language like Python as an abstraction layer to generate circuits is that we are not limited by a drag-and-drop graphical user interface. We have access to the full expressiveness and library support of the language. Graphical Interfaces tend to change in time, while a programming language stays fixed. This eliminates the need of learning different software and users can quickly start designing. What is more, changes in the design are represented as changes in the source code which can make the process of versioning much simpler.

Most of the commercial available software provide an interface to perform parametric analysis on a design. However, if we want to generate different versions of a circuit, each version has to be generated by hand, thus reducing the possible space of exploration due to time or complexity constraints. With our tool, parametric characteristics can be embedded directly into the components and the multiple designs can be generated in a modular and programmable fashion.

\section{Overview of the tool}\label{sec:description}

\subsection{Electrical components and parameters}\label{sec:comp_models}

The parameters of electrical components can be defined statically or dynamically calculated through Python functions or SymPy formulas, that may not be available or as accessible in EDA tools. Dynamic parameters bring the possibility of embedding parametric analysis directly into the circuit definition. These parameters can be grouped into \texttt{ParamSets} that behave similar to process corners. The following example shows a reduced number of parameters for a NMOSFET transistor, where the \textit{vth} of the transistor is drawn from a Gaussian distribution and the \textit{test} parameter is defined by the formula $1 / vth$.

\begin{code}
\begin{minted}{python}
nmos_params = Params(
    {"w": 0.135, "vth": random.gauss(0.4, 0.1), "test": Formula("1 / vth")}
    )
ParamSet({"TT": nmos_params})
\end{minted}
\caption{Example parameters for a NMOSFET transistor where the \textit{vth} is defined dynamically.\label{lst:example_model}}
\end{code}

In order to automatically generate parameters from files that are commonly used, this framework provides a parser interface to extract information from different file formats and PDKs. Multiple parsers are already available but users can extend this functionality by defining their own custom parsers. Certainly this functionality makes the process of testing different technology nodes or constraints more accesible, as the parameters and component names can be can be updated in the moment and swapped in place depending on the desired environment.

Electronic components themselves can be created through the \texttt{Component} class. Besides the basic properties like component name, connected nets and parameters, users can embed metadata to provide additional information that can be shared between different tools. These components serve as templates to generate the modular circuits. Verilog-A components can also be be easily accessed thanks to the provided parser and user created components.

\begin{code}
\begin{minted}{python}
params = Reader.load("/path/to/params_file")

Cap = Component("Cap", [0, 1], params['cap']['TT'], prefix = "C")
model = Model("custom_nmos", "nmos", {"TYPE": 1})
\end{minted}
\caption{Example creation of a capacitor and a custom NMOS model. The parameters are extracted from a file using an example Reader.\label{lst:example_comp}}
\end{code}

Once the components have been defined, it can be instantiated multiple times by using the operators \texttt{@} and \texttt{\%} which are overloaded to quickly modify the connections and the parameters of a component.

\subsection{Manipulations and operations}

As it has been show in section~\ref{sec:state_art}, there are already tools that allow to generate netlists. The core objective of this framework is to provide very efficient manipulation primitives to quickly create complex and reusable connection patterns that can be customised through variables. This framework provides a small list of operations that can be used to create more complex patterns, like the \texttt{Parallel} and \texttt{Chain} operations that create components in parallel and in a daisy chain as their name imply. The manipulations automatically instantiate the number of desired components and update their connections or parameters. In this way, the connection between components occurs in a deterministic and reusable way so it's easier to avoid mistakes when connecting components, which could minimise the need of Electrical Rule Checking (ERC) tools.

\begin{code}
\begin{minted}{python}
NMOS = Component("nmos", [1, "INPUT", 3, "GND"])
Res = Component("res", [1, "GND"])
parallel = Parallel(Res, 3) 
chain = Chain(NMOS, 3, in_port = 0, out_port = 2) 
\end{minted}
\caption{Example of the basic manipulation operations.\label{lst:example_manip}}
\end{code}

Although only a limited number of manipulations are already provided by the library, users can use them to create and extend their own manipulation operations. The components generated by a manipulation can be accessed and modified directly. This ease of modification is handy to simulate process-induced variability or even to evaluate the resilience of a system to faults or errors. Said faults can be injected, as an example, into a list of components and their behaviour can be measured. In the following example, the manipulation \texttt{Inject} receives a chain of components and a probability of defect injection. For each component in the chain it has a chance of generating the desired defect and connecting it to the output of the component. We can also see in this example how the \texttt{Inject} and \texttt{Chain} manipulation can be concatenated to produce the desired circuit.

\begin{code}
\begin{minted}{python}
class Inject(Manip):
    def __init__(self, comps, p = 0.5, defect = None):
        super(Inject, self).__init__()
        defect = defect or Component("Res", ["", "GND"], {"R": 1e4})
        for comp in comps:
            if random.random() <= p:
                # Inject the defect and reconect  
                self.children.append(defect @ [comp.ports[-1], "GND])
            self.children.append(c)

chain_defects = Inject(Chain(mosfet, 7), p = 0.7)
\end{minted}
\caption{Example of defect injection in a chain of 7 transistors.\label{lst:defects}}
\end{code}

Another useful manipulation is the \texttt{Array}, which that allows instantiating components in a 1D or 2D array and their connections can be updated dynamically trough their coordinates, as it is shown in the following example. This array generation utility can be of great use to create crossbar arrays, 2 dimensional CMOS sensors and Micro Electro-Mechanical Systems (MEMS) matrix that contain a very large number of components.

\begin{code}
\begin{minted}{python}
def ports_crossbar(coords):
    x, y = coords
    return [f"X_{x}", f"Y_{y}"]
    
arr_size = (3, 3)
arr = Array(arr_size, device, ports_crossbar)
\end{minted}
\caption{Example of 2D crossbar array. The size of the array is determined by the \texttt{arr\_size} variable.\label{lst:crossbar}}
\end{code}

To allow for reusable designs and more complex logic, multiple components can be grouped inside a \textit{subcircuit}, just like SPICE subcircuits. Subcircuits can be \textit{fixed} so that no more components can be added. This can be used to stop the addition of components in a loop based on logical tests. Once a subcircuit has been defined, it can be used as a component and thus the manipulation primitives can be applied. The components and subcircuits created can be grouped inside a \textit{Circuit}. A circuit behaves very similarly to a SPICE netlist and can be then converted into a subcircuit to be used in other designs. This is the one of the main interfaces for code re-usability and modular designs.

\subsection{Exporting elements}

All the elements created can be exported to text files so that they can be shared between different utilities or read back in a later future. Moreover, this framework provides an interface to export the elements to different file formats that users can extend to create their desired exporters. This process makes the framework simulator agnostic, as the same design can be exported to different simulators just by using different Exporters. Furthermore, users are not only bounded to simulators as the different components and nets can be exported to other kind of file formats for analysis.

\begin{code}
\begin{minted}{python}
exporter = CustomExporter()
exporter.dump(circuit)
# Or directly to a file
exporter.dump_to_file("/path/to/file")
\end{minted}
\caption{Example exporting a design into a file.\label{lst:exporting}}
\end{code}

\subsection{Example of a complex circuit}\label{sec:example}

This tool has been used to create the circuits used in the study~\cite{rel_ro}. A Ring Oscillator is a chain of inverters that oscillates when an input signal is applied to the first inverter in the chain due to the gate delay. Multiple of this Ring oscillators can be connected to multiplexers that allow the selection of a pair of Ring Oscillators. The output signal of the multiplexer can be fed to a counter to measure the oscillation frequency of the Ring oscillators.

\begin{figure}[H]
    \centering
    \includegraphics[width=0.8\columnwidth]{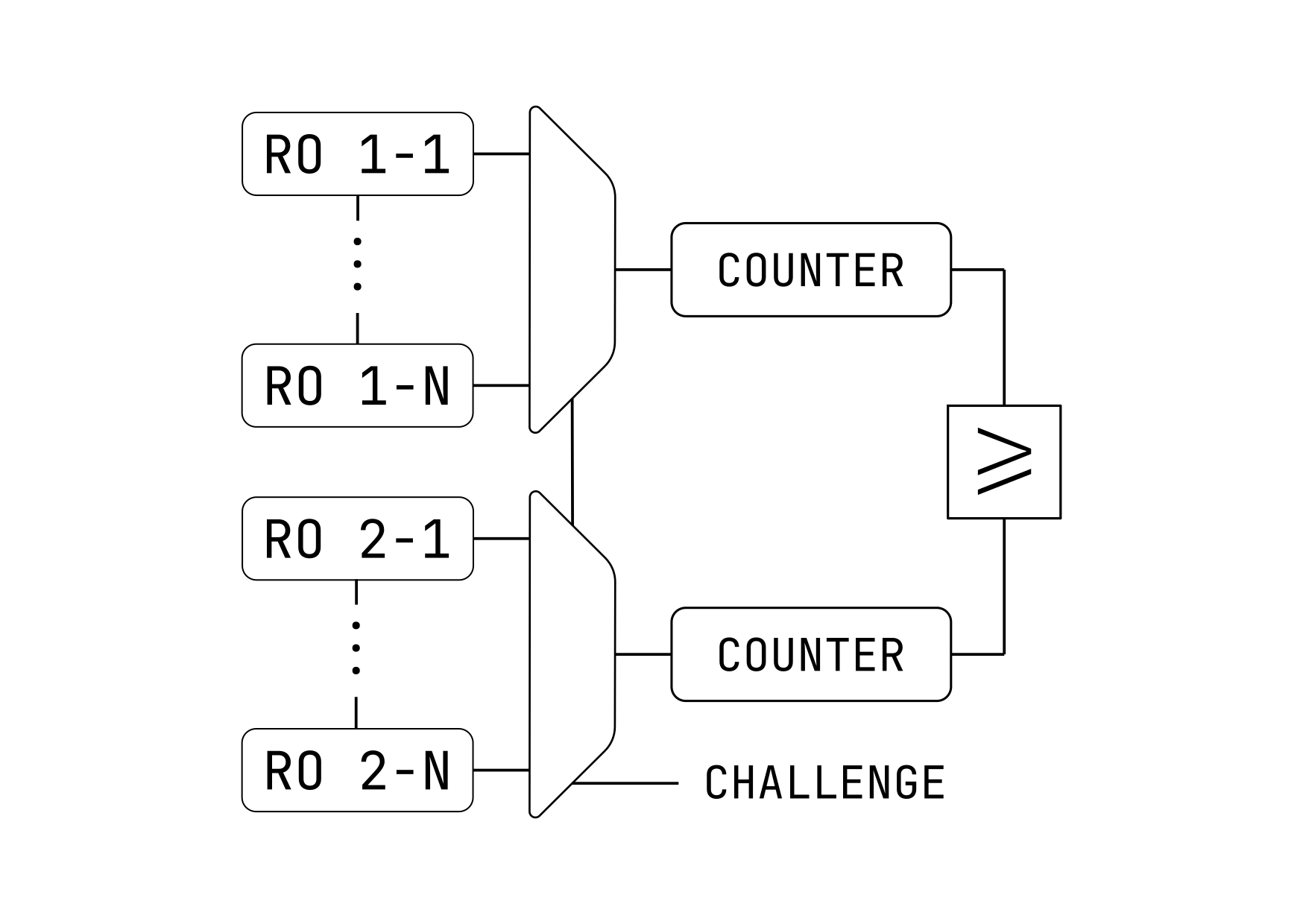}
    \caption{Schematic of a Ring Oscillator Physical Unclonable Function}
    \label{fig:ro_puf}
\end{figure}

In this case the number of inverters per ring oscillator and the total number of chains are determined by the \texttt{N\_RO\_PER\_CHAIN} and \texttt{N\_CHAINS} variables respectively. The number of inputs of the multiplexer can be defined dynamically also from the \texttt{N\_CHAINS} variable. The Counter component has been created in Verilog-A.

\begin{code}
\begin{minted}{python}
reader = VerilogAReader()
VCounter = reader.load("/path/to/counter.va")

# Define the size of the Ring Oscillator
N_RO_PER_CHAIN = 5
N_CHAINS = 3

# Dynamic generation of the multiplexer
MUX = Subcircuit("MUX", [f"IN_{d}" for d in range(N_CHAINS)] + ["Sel", "OUT"], {})

INV = Subcircuit("INV", ["in", "out"], {})
# Components can be added by using the += operator
INV += Mosfet(["out", "in", GND, GND], name="nmos")
INV += Mosfet(["out", "in", VDD, VDD], name="pmos")
inv = INV @ ["in_chain", "1"]

chain = Circuit()
chain += NamedChain(inv, N_RO_PER_CHAIN, out_name="OUT")
ro_chain = chain.into_subckt("RO_CHAIN", ["in_chain", "OUT"], {})

chains = Chain(ro_chain @ ("INPUT", "OUTPUT"), N_CHAINS)
netlist += chains
nodes = []
for comp in chains:
    nodes.append(comp.nodes)

counters = Parallel(VCounter([""]), N_CHAINS)
for i, comp in enumerate(counters):
    comp @= nodes[i][-1]
    
netlist += counters
\end{minted}
\caption{Example of creating the Ring Oscillator Chains.\label{lst:ro_chain}}
\end{code}

\section{Conclusions}\label{sec:conclusions}

The framework shown in this article allows for fast design space exploration and parametrization of electronic designs thanks to its powerful manipulation primitives. It can read components and electrical parameters from a set of file formats and it can export the designs to any major available commercial simulator as well as different output formats specified by the user. 

The next objective of this framework is to allow the automatic management of libraries and layouts as well as the generation of layouts in different formats and technology nodes. Besides, tools like AGS~\cite{asg} and N2S~\cite{n2s} can also supported to create an schematic from the generated netlist.

\bibliographystyle{IEEEtran} 

\end{document}